\documentclass[11pt]{article}
\usepackage{times}
\usepackage{geometry}
\usepackage[utf8]{inputenc}
\usepackage{enumitem,amssymb}
\usepackage{ragged2e}
\newlist{thematic}{itemize}{8}
\setlist[thematic]{label=$\square$}
\usepackage{pifont}
\usepackage{setspace}

\usepackage{fancyhdr}
\usepackage[USenglish]{babel}
\usepackage[utf8]{inputenc}
\usepackage[T1]{fontenc}
\usepackage{epsfig}
\usepackage{wrapfig}
\usepackage{color}

\usepackage[vflt]{floatflt}
\usepackage{longtable}
\usepackage{caption}
\usepackage{jheppub}   
\usepackage[dvipsnames,svgnames,x11names]{xcolor}
\usepackage[all]{hypcap} %Links go to figures; breaks on deluxetables (use \capstartfalse \capstarttrue to fix it)
\usepackage{amssymb,amsmath}
\usepackage{longtable}
\usepackage{amssymb}
\usepackage{amsmath, xspace}
\usepackage{graphics,graphicx} 
\usepackage{rotating}
\usepackage{color}
\usepackage{xcolor}
\usepackage{multirow}
\usepackage{subfigure}
\usepackage{sectsty}
\usepackage[outercaption]{sidecap}
\sidecaptionvpos{figure}{c}
\usepackage{multicol}

\usepackage{natbib}

\let\OLDthebibliography\thebibliography
\renewcommand\thebibliography[1]{
  \OLDthebibliography{#1}
  \vspace{-8px}
  \setlength{\parskip}{0pt}
  \setlength{\itemsep}{0pt plus 0.3ex}
}

\definecolor{DarkGreen}{rgb}{0.0, 0.3, 0.0}
\definecolor{purple}{rgb}{0.5, 0.0, 0.5}
\definecolor{red}{rgb}{1, 0.0, 0.0}
\definecolor{green}{rgb}{0, 1.0, 0.0}

\def\3he{$^3{\rm He}$}

\def\lsim{\mathrel{\lower2.5pt\vbox{\lineskip=0pt\baselineskip=0pt
           \hbox{$<$}\hbox{$\sim$}}}}

\def\gsim{\mathrel{\lower2.5pt\vbox{\lineskip=0pt\baselineskip=0pt
           \hbox{$>$}\hbox{$\sim$}}}}

\begin{document}
\raggedright
\huge
Detection and characterisation of submm transient
\vspace{0.5em}\linebreak
sources with a large single-dish telescope
\linebreak
\bigskip
\normalsize

\textbf{Authors:} \\
%\cc{Note: first 3 authors must be from ESO countries}
Mike Peel (email@mikepeel.net, Imperial College London);\\
Dave Clements (Imperial College London);\\
Tony Mroczkowski (Institute of Space Sciences, ICE-CSIC);\\
Allen Foster (Princeton University)
\linebreak

\textbf{Science Keywords:} % https://www.eso.org/sci/observing/phase1/p117/CfP117.pdf
transients --- submm observations --- surveys
\linebreak

 \captionsetup{labelformat=empty}
\begin{figure}[h]
   \centering
\includegraphics[width=0.7\textwidth, page=3, clip, trim=40mm 40mm 40mm 40mm]{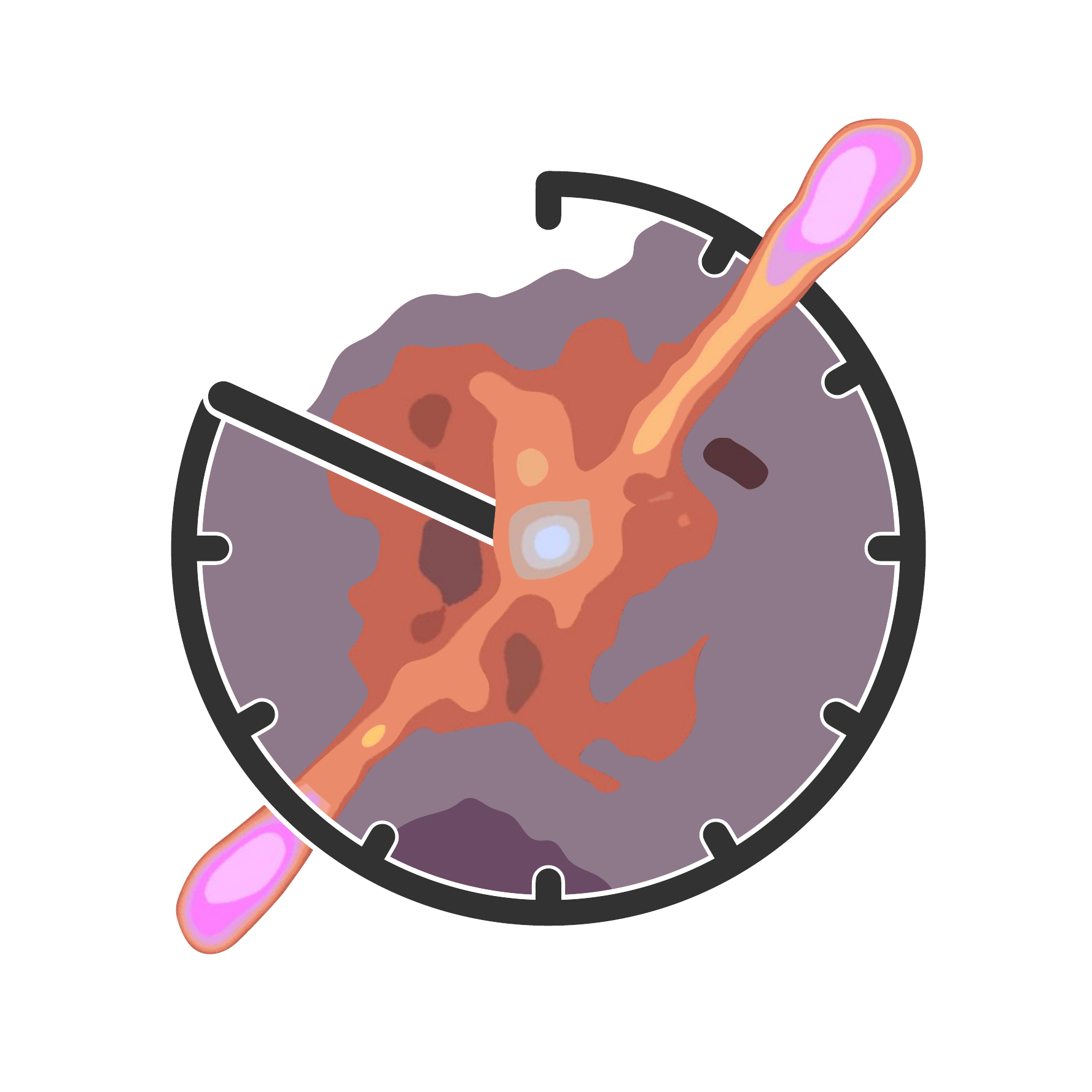}
   \caption{Cartoon depiction of a transient event going off with a clock superimposed to indicate time.  Image from \url{https://atlast-telescope.org/science/} courtesy L. Di~Mascolo.}
\end{figure}
\vspace{-1mm}

\pagebreak
\justifying

\section*{Abstract}
\vspace{-3mm}
The exploration of the time-variable astronomical sky at submm wavelengths is rapidly becoming more feasible with large sky surveys by Cosmic Microwave Background telescopes with tens of thousands of detectors. Observations with the Atacama Cosmology Telescope and South Pole Telescope have already detected some transients, and Simons Observatory and CCAT are expected to detect many more in the near future. Follow-up observations to characterise these transients, and surveying to uncovering fainter populations, will need high sensitivity and large fields of view at submm wavelengths, which could be provided by large single-dish telescopes such as AtLAST.

\vspace{-3mm}
\section{Introduction}
\vspace{-3mm}
Transient sources have long been of significant interest to astronomers, going back to the earliest observations of supernovae. The field is now well developed at optical and radio frequencies, however exploring transients in the submm wavelength regime in between these is still in its early years. Recent observations with Cosmic Microwave Background (CMB) experiments are already yielding new transient detections, in particular the Atacama Cosmology Telescope (ACT) \citep{Li2023,HerivasCaimapo2024} and the South Pole Telescope (SPT) \citep{Guns2021,Tandoi2024}. Simons Observatory (SO) has now started observing and will detect many transients in the near future, particularly with the 6-m Large Aperture Telescope (LAT) \citep{LATASO2025,Clancy2025}.

\vspace{-3mm}
\section{Expected classes of submm transients}
\vspace{-3mm}

Transient submm emission is expected both from sources within and outside of our Galaxy \citep{Clements2025,JackOS2025}. Locally, flaring stars \citep{Tandoi2024} are detected, typically with durations <3 days, and protostars that can have variable accretion rates causing variations on month to year timescales. These are frequently difficult to observe in the optical and infrared, and submm thermal emission---particularly at the shortest wavelengths \cite{Fischer2024}---can provide valuable insights into their evolution. The JCMT Transient Survey of protostars has provided significant information on these in recent years, including the most powerful flare from a young star \citep{JCMT2019,JCMT2024}, with more expected to come from SO surveys soon. Even more locally, objects moving in our Solar System, such as asteroids, are also visible at these wavelengths, and appear as transients since they rapidly move around the sky \citep{Chichura2022,JackOS2024}. These will also be detected by any submm telescope transient search, with higher sensitivities detecting smaller or more distant asteroids.

\begin{figure}[tb]
\begin{center}
\includegraphics[height=12cm]{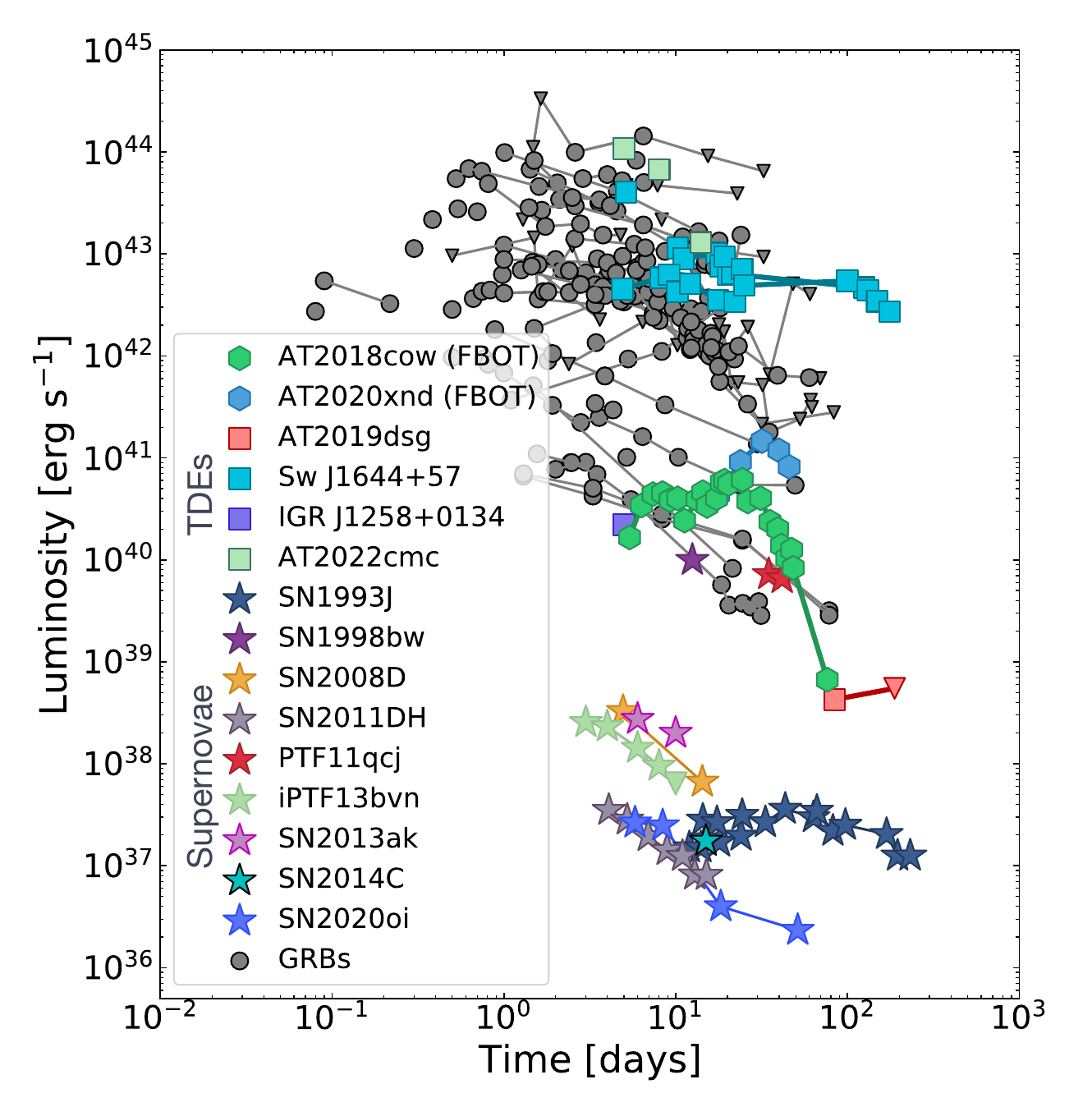}
\end{center}
\vspace{-10mm}
\caption {Light curves of various extragalactic transients (SNe, GRBs, TDEs and FBOTs) at millimetre wavelengths, from \cite{Eftekhari2022}.}
\label{fig:extragalactic}
\end{figure}

The main source of extragalactic transients is expected to be gamma-ray bursts \citep{Eftekhari2022}, along with supernovae (SNe), tidal disruption events (TDEs), and fast blue optical transients (FBOTs), as shown in Fig.~\ref{fig:extragalactic}. FBOTs in particular are from currently unknown origins, and there is particular interest in detecting and characterising more of these. TDEs are also of specific interest, since dust-enshrounded TDEs \citep[e.g.,][]{Mattila2018} can be seen at submm wavelengths but are not so clear in the optical. Variable AGN are also of interest to study over longer periods, and fainter AGN may appear as transients during particularly bright flaring events. Known extragalactic transients have a range of intrinsic luminosities, but are typically visible for weeks to months, making it significantly easier to follow them up.

\vspace{-3mm}
\section{Detection and characterisation possibilities}
\vspace{-3mm}
\textbf{Detections with a new large submm telescope}: Surveys with a single dish telescope, either with multiple epoch observations focused on specific star-formation regions or performing wider sky surveys for other science goals, will detect transient sources directly. This is particularly the case with telescopes with large fields of view and high detector counts, ideally when also coupled with large collecting area. A particular example is the JCMT Transient Survey of protostars \citep{JCMT2024}, which a larger telescope could observe much more deeply or over larger areas of the sky.

\vspace{5px}

\noindent\textbf{Triggered observations after detections with other submm telescopes}: CMB telescopes such as SO and SPT are currently surveying large areas of the sky at mm/submm wavelengths and will yield 10s--100s of transient detections over the next decade. SO's LAT in particular, with its 6\,m collecting area and 60,000 detectors observing at 22--313\,GHz over the next 10 years \cite{LATASO2025}, will find many transient sources, with good observing overlap with a potential new facility in the 2030s or 2040s. The CCAT-prime survey \citep{CCAT2023}, which uses a similar telescope to LAT to observe higher frequencies, will also provide a good source of transient detections. Future expansions of SO, or projects along the lines of the previously planned CMB-S4, would yield even more transient detections each year over a longer time period. Following up these detections with a large single dish telescope would provide additional data points to characterise the spectrum and light curve properties of the transient, especially as it fades post-detection. It will provide more accurate positions of the transients, which will aid with identifying the host objects and follow-up with higher resolution telescopes. Additionally, sources detected by the large telescope can be investigated historically using archival searching of the transient position in SO/SPT/ACT data.

\vspace{5px}

\noindent\textbf{Triggering from other wavelength detections}: Some transients detected at other wavelengths, for example in the optical with the Rubin Legacy Survey of Space and Time \citep{LSST2009}, or radio surveys with ASKAP, MeerKAT, and the Square Kilometre Array (SKA), will also have submm counterparts. Routine follow-up of particular classes of these transients, based on brightness and spectral cuts that can be defined using Rubin's transient brokers, can be used to statistically characterise their submm properties. A large single dish submm telescope can quickly follow up such transients, while CMB telescopes will mostly get coincident observations when transient sources are within their standard survey observations, and instruments like ALMA are typically very over-subscribed and limited to follow-up in one band at a time. It can also follow up non-electromagnetic transient sources such as gravitational wave events (e.g., from LIGO, LISA, and/or the Einstein Telescope), where a larger region of the sky needs to be surveyed for candidate electromagnetic counterparts.

\vspace{-3mm}
\section{Technical requirements}
\vspace{-3mm}
The ideal instrument for this work is a large ($\sim$50\,m) single dish telescope that can simultaneously observe in the frequency range $\sim$30--950\,GHz with a large field of view and focal plane area, located at high altitude to minimise atmospheric noise, and with flexible scheduling to enable rapid follow-up of transient detections. This would be highly complementary to Simons Observatory and ALMA observations at similar frequencies. The Large Millimetre Telescope (LMT) \citep{Hughes2020} meets some of these requirements, particularly with aperture and partial frequency coverage. However the AtLAST proposal \citep{Mroczkowski2025}, with its 50\,m primary mirror, large 2$^\circ$ field of view (versus LMT's 4' FoV), fast mapping speed, and wide frequency coverage, is particularly well suited to fulfilling these needs \citep{JackOS2025}.

A potential issue is that submm telescopes will also see low earth orbit satellites through their thermal emission \citep{Foster2025}, no matter where they are located on Earth, which can be confused with transients unless the satellite positions are well known \citep{Peel2025}. Given the rapidly increasing number of satellite constellations, the sooner submm observations can be performed, the better, particularly as it is not (yet) practical to perform similar observations with a $\sim$50\,m submm telescope from orbit or from the far side of the Moon.

\vspace{-3mm}
\section{Conclusions}
\vspace{-3mm}
A $\sim$50\,m submm telescope with a large field of view, many detectors, multiple simultaneous bands, and fast survey capacity (such as that proposed by AtLAST) could provide numerous detections of submm transient sources, and would also help improve localisation and characterise transients detected by other telescopes. It would provide complementary information on transients to that from existing facilities such as ALMA and SO, and would also connect with proposed observations with infrared telescopes such as PRIMA \citep{Clements2025}, alongside accomplishing many other important scientific goals.

%\vspace{-3mm}
\begin{multicols}{2}
\bibliographystyle{aasjournal}
\bibliography{references.bib}{}
\end{multicols}

\end{document}